\documentclass[twocolumn]{aastex631}

\usepackage{graphicx}% Include figure files
\usepackage{dcolumn}% Align table columns on decimal point
\usepackage{bm}% bold math
\usepackage[nolist,nohyperlinks]{acronym}
\usepackage{hyperref}% add hypertext capabilities
\usepackage{color}
\usepackage{xspace}
\usepackage{subfigure}
\usepackage{amsmath}
\usepackage[normalem]{ulem} %for \sout
\usepackage{hyperref}

\newcommand{\fx}[1]{{\color{black}{{#1}}}}

\newcommand{\cg}[1]{{\color{black}{{#1}}}}

\begin{document}

\title{Inference of multi-channel r-process element enrichment in the Milky Way \\ using binary neutron star merger observations}

\author[0000-0001-5403-3762]{Hsin-Yu Chen}
\email{hsinyu@austin.utexas.edu}
\affiliation{Department of Physics, University of Texas at Austin, Austin, Texas 78712, USA}
\author[0000-0002-8457-1964]{Philippe Landry}
\email{plandry@cita.utoronto.ca}
\affiliation{Canadian Institute for Theoretical Astrophysics, University of Toronto, Toronto, ON M5S 3H8, Canada}
\author[0000-0002-3923-1055]{Jocelyn S. Read}
\email{jread@fullerton.edu}
\affiliation{Department of Physics, California State University Fullerton, Fullerton, CA, USA}
\author[0000-0001-6374-6465]{Daniel M. Siegel}
\email{daniel.siegel@uni-greifswald.de}
\affiliation{Institute of Physics, University of Greifswald, D-17489 Greifswald, Germany}
\affiliation{Department of Physics, University of Guelph, Guelph, Ontario N1G 2W1, Canada}

\begin{abstract}
Observations of GW170817 strongly suggest that binary neutron star (BNS) mergers produce rapid neutron-capture nucleosynthesis (r-process) elements. However, it remains an open question whether these mergers can account for all the r-process element enrichment in the Milky Way's history. 
Here, we constrain the contributions of the BNS channel using astrophysical neutron star observations. The rate and mass distributions are constrained by LIGO/Virgo/Kagra through the latest catalog GWTC-3, the neutron star equation of state by gravitational-wave, radio, and X-ray observations, and the delay time distribution by short gamma-ray burst (GRB) host galaxy associations. We present a Bayesian framework to consistently combine {these  observations with abundance information to quantify the contribution and uncertainties of single and multiple astrophysical enrichment sources, and obtain a distribution of per-event BNS r-process element yields consistent with geophysical and astrophysical abundance constraints}.
%In a one-zone Galactic chemical evolution model, the resulting distribution of per-event r-process element yields is consistent with geophysical and astrophysical abundance constraints.}
%{However, it is insufficient to explain the relative r-process abundances of disk stars in the Galaxy. 
{We then adopt a galactic chemical evolution model assuming instantaneous and fixed amount of Fe enrichment from core-collapse supernovae, and show that BNS-only enrichment scenarios remain inconsistent with the observed r-process abundance trend of disk stars in the Galaxy even with the uncertainties in BNS merger observations.}
Using stellar abundance observations instead of the short GRB constraints, we can infer a shorter BNS delay time distribution with power-law index \fx{$\alpha\leq -2.0$} and minimum delay time \fx{$t_{\rm min}\leq 40$ Myr} at 90\% confidence{, consistent with detailed galactic chemical evolution models.}
%We then use a one-zone Galactic chemical evolution model to show that BNS would have to merge shortly after the formation of the stars, with a delay time distribution parameters of \fx{$t_{\rm min}\leq0.043$ Gyr} and power-law index \fx{$\alpha\leq -1.98$}, in order to explain the Galactic stellar observations in the SAGA database. 
Such delay times are in tension with those predicted by standard BNS formation models.
{Alternatively, we confirm that a two-channel scenario, in which the second channel tracks the star formation history without significant delay, can account for both Galactic stellar and short GRB observations.} 
%We quantitatively infer properties of the second channel from astronomical observations.}
%Instead, a multi-channel enrichment scenario can account for the Galactic stellar observations with the second channel closely following the star formation and contributing to \fx{at least 40\%} of r-process elements at $z=0$.
%Our results suggest that \fx{$64^{+0.23}_{-0.28}\%$} of the r-process abundance in the Milky Way today was produced by extra channels other than BNSs, or a secondary BNS merger population. 
{We estimate that \fx{45--90\%} of the r-process abundance in the Milky Way today {would have been} produced by this star-formation-tracking channel, rather than BNS mergers with significant delay times.}
\end{abstract}

%\maketitle
\section{Introduction}

A fundamental question of nuclear astrophysics is the origin of the elements in the solar system, the Milky Way, and the Universe as a whole. While it is %relatively well-
understood how elements up to iron are forged in the cores of stars and ejected through supernovae to enrich the metallicity of future stellar generations, the origins of the rapid neutron-capture (`r-process') elements, which must be produced in dense neutron-rich environments, remains open to active discussion \citep{cowan_origin_2021,siegel_r-process_2022,arcones_origin_2022}. 

Observational constraints from various astrophysical environments tell us the characteristics of the events that contribute r-process elements to our Universe. Geophysical constraints \citep{wallner_abundance_2015,hotokezaka_short-lived_2015} as well as observed levels and scatter of r-process elements in metal-poor stars in different environments, such as the Galactic halo \citep{macias_stringent_2018}, ultra-faint dwarf galaxies \citep{ji_r-process_2016,beniamini_r-process_2016}, and globular clusters \citep{roederer_heavy-element_2011,roederer_detailed_2016}, suggest that they must be rare but prolific events in both recent and early Galactic history%. For a summary and discussion of rate-yield constraints, 
---see, e.g.,~\citet{hotokezaka_neutron_2018} and \citet{siegel_gw170817_2019}. Observed abundances of r-process tracer elements such as Europium relative to stellar metallicity over the history of iron enrichment by supernovae, from the metal-poor stars of the Galactic halo to the high-metallicity environments of Galactic disk stars, suggest a significant %, if not dominant,
fraction of Galactic r-process events must have a short delay time with respect to star formation \citep{2014MNRAS.438.2177M,2015MNRAS.452.1970W,hotokezaka_neutron_2018,siegel_gw170817_2019,SiegelBarnes2019,cote_neutron_2019,2020ApJ...900..179K,lian_observational_2023,van_der_swaelmen_gaia_2023}, although this has been questioned \citep{banerjee_neutron_2020,tarumi_evidence_2021}. Furthermore, the brief star formation histories in (ultra-faint) dwarf galaxies of only $\mathcal{O}(100\,\text{Myr})$ and of $\mathcal{O}(10\,\text{Myr})$ in globular clusters also require very short delay times for internal r-process enrichment \citep{ji_r-process_2016,skuladottir_neutron-capture_2019,naidu_evidence_2022,zevin_can_2019,kirby_stars_2020,kirby_r-process_2023}.

%[WHY THIS WORK IS NEEDED]
However, when evaluating candidate channels for r-process contributions in various environments, it has often been assumed that the overall rate of events, the amount of r-process element-containing ejected matter from each event (the `yield'), and the delay with respect to star formation are largely unconstrained and can be tuned to the levels required by abundance observations. In addition, previous work exploring multi-channel r-process enrichment~\citep{2014MNRAS.438.2177M,2015MNRAS.452.1970W,cote_neutron_2019,2020ApJ...900..179K,2023ApJ...943L..12K} focused on point estimates for fixed scenarios{, such as a fixed binary neutron star (BNS) merger rate or a fixed fractional secondary channel contribution. The advent of gravitational-wave (GW) astronomy establishing BNS mergers as one site for r-process nucleosynthesis via GW170817, the subsequently improving constraints on merging neutron star population properties~\citep{abbott_gw170817_2017}, on the equation of state of NSs as well as on the delay time distribution of sGRBs provide a wide array of empirical information that strongly affects the modeling of chemical evolution of r-process elements.} 
%With the advent of gravitational-wave (GW) astronomy %establishing BNS mergers as one site
%(potentially among several) 
%for r-process nucleosynthesis via GW170817, and with subsequently improving constraints on merging neutron star population properties~\citep{abbott_gw170817_2017}, this is no longer true for neutron star mergers. 
{In this work,} we bring together several {new} lines of observational evidence {and develop a Bayesian inference framework to appropriately take into account the uncertainties in different observational products} to quantitatively constrain the contribution of BNS merger events to r-process nucleosynthesis. 

First, we make use of direct constraints on the rates and properties of BNS mergers from LIGO-Virgo-KAGRA (LVK) observations over the first three observing runs \citep{the_ligo_scientific_collaboration_gwtc-3_2021,KAGRA:2021duu}. Past results (e.g.,~\citet{hotokezaka_neutron_2018,siegel_gw170817_2019,cote_neutron_2019}) focus on the rate estimates made immediately after the first observation of a BNS merger GW170817~\citep{abbott_gw170817_2017}, which assumed that all BNS mergers have similar properties. Additional events, and (thanks to a well-modeled sensitive volume over the following observing runs) observing time without events, however, furthered our understanding of the merger population as a whole. GW astronomy is now producing rate estimates as a function of component masses, which extend beyond the point-observation of GW170817 \citep{LIGOScientific:2018hze}.

{Second}, we take a data-driven posterior distribution for the neutron-star EOS at high density, which combines information from an array of astronomical observations of neutron stars. Here, we use the public EOS posterior samples of \citet{2021PhRvD.104f3003L}, which are informed by pulsar mass measurements, GW observations, and X-ray pulse profile modeling. The EOS constraints encoded in this posterior distribution are broadly consistent with those of e.g.~\citet{2022Natur.606..276H,2020NatAs...4..625C}, which include additional astrophysical and nuclear-physics constraints.

{Third}, we take an empirical estimate for the delay time distribution of neutron-star mergers that comes from the observations of the galaxy offsets of short gamma-ray bursts (sGRBs)~\citep{ZevinNugent2022}. This provides limits on the distribution of delay times for merger events, assuming that sGRB associations are an unbiased probe of the merging BNS population. Notably, the power-law slope of the sGRB-inferred delay time distribution is steeper than the conventional expectation of $\alpha \approx -1$ for BNSs formed through isolated binary evolution~\citep{Piran1992,DominikBelczynski2012,MapelliGiacobbo2018,NeijsselVignaGomez2019}.

%{We develop a Bayesian inference framework to appropriately take into account the uncertainties in different observational products and place quantitative constraints the amount of r-process elements contributed by BNS mergers.}
%\hy{Move the following to discussion?} \PL{[Personally, I like having the summary here and would move the blue text above to L80.]} \hy{I also like this summary, but the referee is asking to shorten the introduction.}\ds{I like the blue sentence above and Phil's suggestion of moving it up to ~L80. We could even sharpen the novel aspect of our work by directly contrasting it with previous work, something along the lines of ``Whereas previous work exploring multi-channel r-process enrichment (citations) focused on point-estimates for fixed scenarios, here we develop a Bayesian...''}\ds{I also like the conceptual summary below, but to satisfy the referee, we may want to try to shorten this to just a few lines and/or move some of this to the discussion?}
Together, these {new observations} have quantitative implications for the amount of r-process elements that will be contributed by BNS mergers{. We introduce our methods and present the four key results in Sections~\ref{sec:rateyield} to ~\ref{sec:multi}. We then summarize and discuss the results in Section~\ref{sec:discussion}.}

\section{Binary neutron star r-process production rate and yield} \label{sec:rateyield}

The r-process enrichment event rate and per-event yield have been constrained by astrophysical and geological measurements. In order to test whether BNS mergers are consistent with these constraints, we use LVK observations to estimate the BNS merger rate, and analytical fits to numerical simulations to estimate the average amount of r-process ejecta per merger event.  
\begin{itemize}
\item\textit{Local merger rate:} We use the BNS population distribution inferred from the LVK's GWTC-3 catalog {for the local ($z<0.1$) merger rate}~\citep{the_ligo_scientific_collaboration_gwtc-3_2021}. Specifically, we adopt the \texttt{Power Law + Dip + Break} population model from~\citet{KAGRA:2021duu}; other population model choices in that study predict consistent merger rates.
%but we do not expect a different choice of population model to affect our results in a significant way. 
\citet{KAGRA:2021duu} provides 1501 posterior rate samples binned in mass. For every rate sample, we sum the rate density over mass bins up to the maximum mass for a given neutron star EOS sample and convert the resulting BNS merger rate to a Galactic event rate, $R_{\rm MW}$, assuming a number density of Milky-Way like galaxies of 0.01 Mpc$^{-3}${~\citep{2016ApJ...820..136G}}. %for direct comparison with other observations.

\item\textit{Neutron star equation of state:} The amount of ejecta produced by a BNS merger depends sensitively on the neutron star EOS. Although the high-density EOS is still uncertain, astronomical observations and recent developments based on chiral effective field theory and many-body perturbation theory in nuclear theory (e.g., \citealt{drischler_how_2020}) have started to constrain and quantify the uncertainties. \citet{2021PhRvD.104f3003L} conditioned a phenomenological Gaussian process model for the EOS on X-ray, radio, and GW observations of neutron stars to provide a set of posterior EOS samples and corresponding neutron star mass-radius relations. For every merger rate sample, we randomly pick a mass-radius relation sample from this set to map from neutron star component masses $(m_1,m_2)$ to radii $(R_1,R_2)$. %We then use the $(m_1,m_2)$ and $(R_1,R_2)$ association to estimate the ejecta mass for a given BNS merger.

\item\textit{Ejecta mass estimate:} We estimate ejecta mass per event, $m_{\rm ej}$, following the approach in ~\citet{2021ApJ...920L...3C}: 
\begin{equation}
m_{\rm ej}=\alpha_{\rm dyn}m_{\rm dyn}+f_{\rm loss}m_{\rm disk},
\end{equation}
where $m_{\rm{dyn}}$ represents the mass of the dynamical ejecta, and $m_{\rm{disk}}$ the mass of the disk formed in the postmerger phase. In order to estimate $m_{\rm{dyn}}$ and $m_{\rm{disk}}$, we use analytical fits to numerical simulations in~\citet{dyn}---specifically, their Eq.~(6) for the dynamical ejecta and Eq.~(4) for the disk---which depend on the mass and radius of the component neutron stars. The coefficient $\alpha_{\rm{dyn}}$ is a scaling factor which is randomly sampled between $[0.5,1.5]$ to account for an uncertainty of 50\% in the knowledge of $m_{\rm{dyn}}$. The fraction $f_{\rm{loss}}$ of mass ejected from the disk is randomly sampled between $[0.15,1]$ to account for its uncertainty. {Both of these uncertainties are taken from the variations in existing numerical simulations of neutron-star mergers~\citep{2021ApJ...920L...3C}.}
%\ds{I wonder whether random sampling of $\alpha_{\rm dyn}$ and $f_{\rm loss}$ favors extreme corner cases. I wonder whether e.g. Gaussian distributions around $\alpha_{\rm dyn}=1.0$ and $f_{\rm loss}=0.3$ with some variance would make a difference in terms of r-process yields?}
\end{itemize}

\begin{figure}
    \centering
    \includegraphics[width=0.95\linewidth]{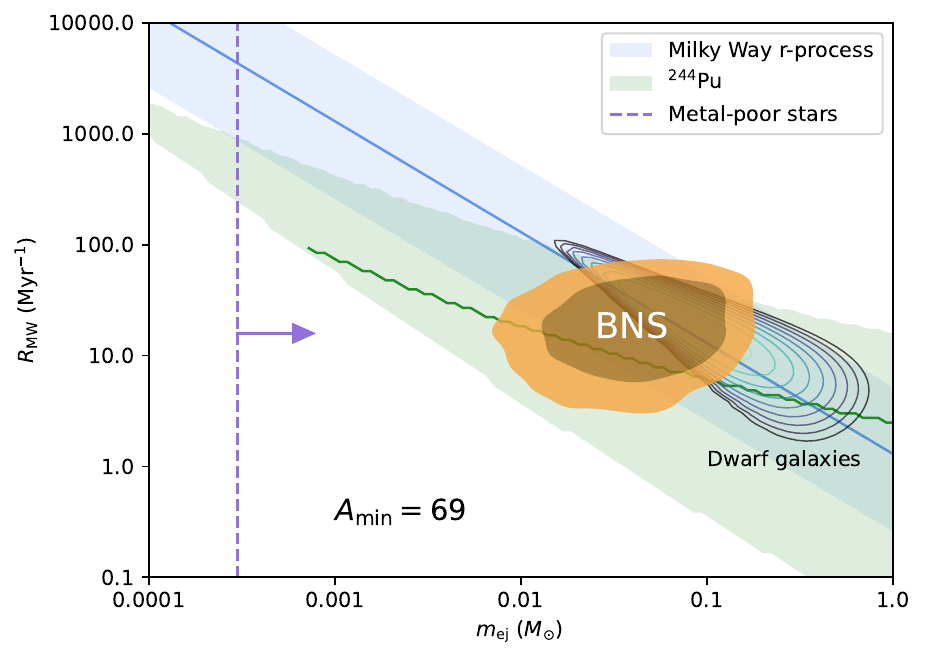}
    \caption{BNS event rate per Milky Way like galaxy as a function of average ejecta mass per event. The dark and light orange shades are the 68\% and 90\% confidence intervals. We over plot the constraints from total stellar mass in Milky Way (~\citet{2011MNRAS.414.2446M}, blue band), metal-poor stars (~\citet{2018ApJ...860...89M}, purple, lower limit), dwarf galaxies (~\citet{2016ApJ...832..149B}, grey contours), and $^{244}$Pu measurements (~\citet{2015NatPh..11.1042H}, green band). We assume an r-process with the solar abundance pattern for atomic mass number $A\geq 69$.}
    \label{fig:prodrate69}
\end{figure}

Given these three prescriptions, for each rate and EOS sample, we sample uniformly in component masses $(m_1,m_2)$ up to the maximum neutron star mass, estimate the ejecta mass for each of these BNS realizations, and perform a rate-density-weighted sum over BNS realizations to obtain the average ejecta mass per event.
%For each rate sample, we estimate the total ejecta mass for every $(m_1,m_2)$ pair up to the maximum neutron star mass, weight them by the corresponding rate density, and sum over all mass pairs to obtain the average ejecta mass per BNS event. 
In Fig.~\ref{fig:prodrate69}, we show the inferred BNS merger rate as a function of average ejecta mass per event. 

We compare our results to different geographical and astrophysical constraints summarized in ~\citet{hotokezaka_neutron_2018}, and find them consistent with {the following constraints from observations:} 
\begin{itemize}
    \item \textit{The total galactic r-process mass}: Assuming the stellar mass in the Milky Way of $6.4\times 10^{10}\,M_\odot$~\citep{2011MNRAS.414.2446M} follows a solar abundance pattern {and integrating the mass of elements with atomic number $A\geq 69$ to obtain the total mass of r-process elements. The average ejecta mass per event is then a function of r-process event rate assuming the age of the Milky Way is $\sim 10$ Gyr.
    \item \textit{Lower limit on the r-process yield per event}: The limit comes from abundance observations of metal-poor stars that have presumably been enriched by a single event only~\citep{2018ApJ...860...89M}.
    \item \textit{Dwarf galaxies}:  The difference between the observations of strongly disparate r-process enrichment in dwarf and ultra-faint dwarf galaxies indicates the rarity of r-process events, which can be converted into rate~\citep{2016ApJ...832..149B}}.
    \item \textit{Geophysical $^{244}$Pu measurements in the deep sea crust and meteorites}: The deep sea crust measurements represent r-process deposition over the past tens of Myrs, and those from meteorites indicate the abundance $\approx\!4.5$\,Gyr ago. Their total abundance and the comparison between the two jointly constrain the rate and yield of r-process events~\citep{2015NatPh..11.1042H}
\end{itemize}
 %\hy{here to add more details about other studies} %\hy{shall we use DTD to obtain the average rate across Milky Way lifetime?}.
%\ds{Whereas the Pu constraints apply to recent Galactic history, the constraints from dwarf galaxies and metal poor stars apply to early galactic history. Therefore, working with average rates over the lifetime of the MW may be a compromise. However, the difference in local to average rates for BNS should only be a factor $\approx\!2$, smaller than or comparable to other uncertainties, so this won't change the figure/conclusions much. One important conclusion from these constraints is that the main r-process site is of rare, high-yield nature *both* in early and recent Galactic history.}

{We note that various assumptions had to be made to produce the geographical and astrophysical constraints summarized above (see ~\citet{hotokezaka_neutron_2018} for more details). The consistency between our results only suggests that BNS mergers \textit{can} explain the rate and yield derived from other observations, but it does not exclude the possibility of other r-process enrichment channels.}

%\begin{figure}
%    \centering
%    \includegraphics[width=0.95\linewidth]{mej_gal_lcehl_nicer_numuncertainty_r90.pdf}
%    \caption{\hy{Is there a value to show this plot as well?} \ds{This is useful if one wants to focus on the main r-process only. Potentially several other sources can contribute to light r-process elements (it is argued in the literature that the observed large scatter among light r-process elements among halo stars is indicative of this)}}
%    \label{fig:prodrate90}
%\end{figure}

\section{Chemical abundance evolution}

Although the BNS merger rate and average ejecta mass is consistent with existing constraints at {high metallicity, where $z\simeq0$,}  a different picture may emerge when enrichment is viewed as a function of cosmic time in various environments and in the context of chemical evolution including other metals. The observed abundances of r-process tracer elements like Eu as a function of stellar metallicity [Fe/H] record the r-process element enrichment at different stages of Milky Way evolution~\citep{2021A&ARv..29....5M}. The r-process enrichment history associated with the BNS merger channel depends on its rate evolution, which can be expressed in terms of a star formation rate convolved with a distribution of delay times measuring the lag between the formation of the progenitor system and the merger. In order to examine whether BNS mergers can produce the measured stellar abundances, we use sGRB observations to inform the BNS delay time distribution and adopt a {simplified} one-zone model {from \citet{SiegelBarnes2019}} to estimate chemical abundances over time.

\begin{itemize}
 
\item \textit{Delay time distribution from short gamma-ray bursts:} 
The BNS merger rate inferred from GWs is confined to the local Universe, and the evolution of the merger rate across cosmic history is currently beyond the sensitivities of the LVK observatories. However, observations of EM counterparts of BNS mergers, such as sGRBs, can provide insight into the merger rate at higher redshift, assuming they appropriately sample the merging BNS distribution. Using 68 identified host galaxies of sGRBs, \citet{ZevinNugent2022} constrained the sGRB rate evolution. Specifically, the rate of sGRBs at cosmic age $t$ can be written as
\begin{equation}
    \begin{aligned}\label{eq:rate}
    R_{\rm sGRB}(t)& = \\
    C_{\rm sGRB}& \int_0^t D(t-t'|t_{\rm min},t_{\rm max},\alpha)\Psi_{\rm SFH}(t')dt',
\end{aligned}
\end{equation}
where $\Psi_{\rm SFH}(t)$ represents the Galactic star formation rate per unit time; we assume the Galactic star formation history follows the cosmic star formation history~\citep{2017ApJ...840...39M}. 
The normalization factor $C_{\rm sGRB}$ {represents} the observed local sGRB rate. The delay time distribution (DTD)
\begin{equation}
    D(t|t_{\rm min},t_{\rm max},\alpha)\propto t^{\alpha}, \, t_{\rm min}\leq t \leq t_{\rm max} \label{eq:DTD_power-law}
\end{equation}
 describes the time between the formation of the progenitor stars and the onset of sGRBs. \citet{ZevinNugent2022} found a relatively steep power-law index of $\alpha=-1.83^{+0.35}_{-0.39}$ and a long minimum delay time of $t_{\rm min}=184^{+67}_{-79}$ Myr. No strong constraint could be placed on $t_{\rm max}$. We assume $t_{\rm max}$ to be the current age of the Universe. While this is not a strong assumption for our modelling, we do know of BNS systems that have a longer inspiral time than the age of the Universe. Although it is uncertain whether sGRBs are representative of the entire population of BNS mergers, the sGRB population is still one of the best indicators of the evolution of the BNS merger rate. Therefore, we use the local BNS merger rate $C_{\rm BNS}$ inferred from LVK observations~\citep{KAGRA:2021duu} to calibrate Eq.~\eqref{eq:rate} instead of $C_{\rm sGRB}$ and project the BNS rate $R_{\rm BNS}(t)$ over time.

\item \textit{One-zone model:} We describe the Galactic production of iron-peak and r-process elements over cosmic time using the one-zone chemical evolution model of~\citet{SiegelBarnes2019}. One-zone models assume instantaneous homogeneous mixing of metals in the interstellar medium (ISM), and thus cannot adequately capture effects of hierarchical structure growth and inhomogeneities due to incomplete mixing associated with rare enrichment events at low metallicity. However, they can accurately capture the average enrichment levels of the ISM with r-process elements at high metallicity, once the ISM has been polluted by a significant number of individual events. %\hy{Daniel, could you elaborate a bit about iron here as the Referee request?} 
We use the same Galactic in- and outflow rate and the same {observationally inferred} rates and yields for iron production by core-collapse and Type Ia supernovae as in~\citet{SiegelBarnes2019}. {Iron-producing core-collapse supernovae are assumed to follow star formation without delay with a local rate of $7.05\times10^{-5}$\,Mpc$^{-3}$\,yr$^{-1}$ \citep{Li2011} and a yield of 0.074\,$M_\odot$ per event \citep{maoz_star_2017}. Type Ia supernovae produce 0.7\,$M_\odot$ iron per event and, consistent with that assumption, they are assumed to follow the star formation history with a DTD of the form Eq.~\eqref{eq:DTD_power-law} with $\alpha=-1.0$ and $t_{\rm min} = 40$\,Myr \citep{maoz_star_2017}, calibrated to a production efficiency of $1.3 \times 10^{-3}$ per $M_\odot$ of stellar mass formed \citep{maoz_star_2017}. We} adjust the rate and yield of r-process enrichment as traced by Eu according to the local merger rate, DTD, and ejecta mass estimates discussed above. The total r-process ejecta mass per event is translated into a Eu mass per event by assuming a solar abundance distribution of r-process elements~\citep{Arnould2007} starting at atomic mass number $A=69$. Representing the cumulative total over many individual enrichment events and a rough average to the r-process abundance patterns of many individual observations of r-process enhanced metal-poor stars that have been polluted by a single or only by a few r-process events (e.g.~\citealt{frebel_observations_2023}), the solar r-process abundance pattern provides a reasonable estimate for average r-process enrichment at high metallicity. {We note that this assumption was made due to the limit of detailed r-process abundance pattern from BNS mergers, and deviations from the solar pattern have been shown~\citep{2023ApJ...943L..12K}.}
\end{itemize}

\begin{figure}
    \centering
    \includegraphics[width=1.0\linewidth]{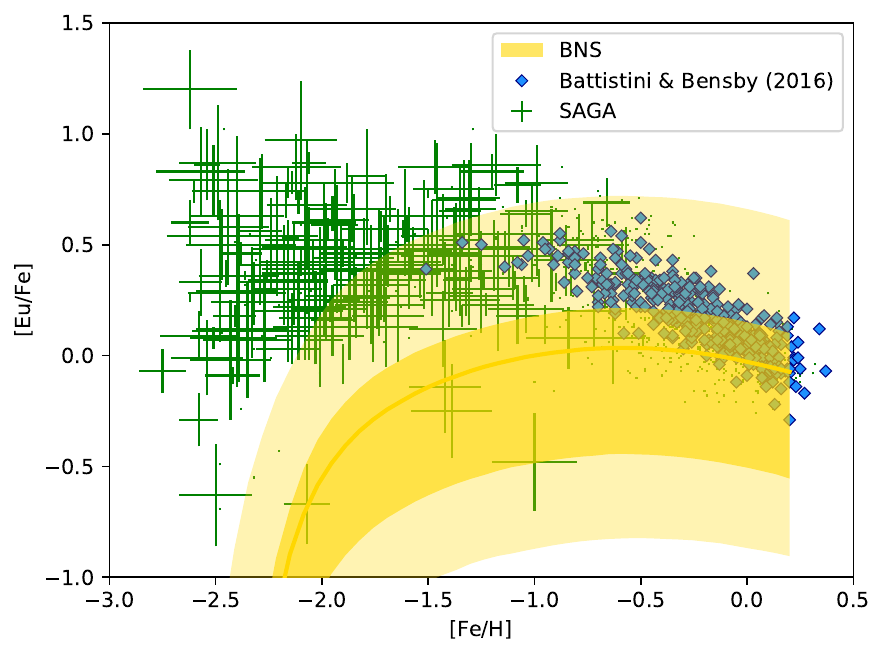}
    \caption{The abundance ratio [Eu/Fe] as a function of metallicity [Fe/H] assuming BNS mergers as the only r-process element production site. The yellow line is the median. The dark and light bands are the 68\% and 90\% confidence intervals. We also overplot the Galactic stellar observations taken from~\citet{BattistiniBensby2016} (blue diamonds) and the SAGA database (\citet{saga}; green crosses).
    }
    \label{fig:FeH_EuFe}
\end{figure}

We first consider an r-process enrichment scenario in which BNS mergers with an sGRB-informed DTD are the only enrichment site. Fig.~\ref{fig:FeH_EuFe} shows the abundance ratio of [Eu/Fe] as a function of [Fe/H]. We overplot the observed metallicities of Galactic disk stars from~\citet{BattistiniBensby2016} (blue diamonds), as well as observations from the Stellar Abundances for Galactic Archaeology (SAGA) database (\citet{saga}; green crosses), which span both metal-rich disk stars and metal-poor halo stars. Despite the large statistical uncertainty originating from the uncertainties in BNS merger rate (locally and at higher redshift), mass distribution, neutron star EOS, and analytic fits to ejecta in numerical simulations as described in the previous section, the abundance ratio [Eu/Fe] estimated from the BNS-only model shows a clear deviation from the stellar observations at low metallicity{, consistent with what was found in previous work.} Thus, while this single-site model {can match the average observed Eu abundance at solar metallicity,}
%can match the observed Eu abundance at $z=0$
 it cannot reproduce the Galactic history of r-process enrichment.

\section{Binary neutron star delay times inferred from Milky Way disk stars}

We showed above that BNS mergers alone cannot match the Galactic stellar observations that probe Milky Way chemical abundance history, given reasonable assumptions about their astrophysical rate, population, delay time distribution and EOS. We now relax these astrophysical assumptions and ask what it would take for BNS mergers to reproduce the history of r-process chemical evolution in the Galaxy. Fig.~\ref{fig:FeH_EuFe} shows that the median of the predicted [Eu/Fe] abundance ratio is consistent with stellar observations around solar metallicity ([Fe/H]$\gtrsim 0.0$), but that the deviation grows with decreasing metallicity. Among the sources of uncertainty, the local merger rate, mass distribution, and neutron star EOS affect the overall abundances of r-process elements, i.e., the vertical offset in Fig.~\ref{fig:FeH_EuFe}. On the other hand, the slope of [Eu/Fe] as a function of [Fe/H] is dominated by the merger rate evolution over time. Therefore, instead of using the delay time distribution informed by sGRBs, we try to determine the BNS DTD that could explain the observed trend of r-process abundances.

We model each stellar measurement of [Eu/Fe] and [Fe/H] as a two-dimensional Gaussian likelihood function $P(d_i | {\rm [Eu/Fe]}, {\rm [Fe/H]})$ with zero covariance. Our results are conditioned on the~\citet{BattistiniBensby2016} disk stars only, as the one-zone chemical evolution model is strictly valid solely at high metallicities when the interstellar medium has already been enriched by many r-process events and there exist well-defined average r-process abundances. The width of the Gaussian likelihood function is dictated by the average standard deviations quoted in~\citet{BattistiniBensby2016}. %For the SAGA data, we adopt the standard deviations that are reported for the individual observations in~\citet{saga}; for data points without reported errors in both [Eu/Fe] and [Fe/H], we assign the average uncertainties from the rest of the observations. 

Within the one-zone model connecting a BNS population to an r-process abundance history, we sample randomly in local ($z=0$) BNS rate-yield ($m_{\rm ej}R_\mathrm{MW}$) and DTD parameters ($t_{\rm min}$,$\alpha$) and calculate the corresponding [Eu/Fe] vs [Fe/H] track. All other population parameter uncertainties---such as those in the NS mass distribution and EOS---are implicitly marginalized over in determining the distribution of possible BNS merger rates and yields. The likelihood of the local BNS rate-yield and DTD parameters is then given by
\begin{equation}
    \begin{aligned} \label{dtd_like}
    &P(d | t_{\rm min},\alpha,m_{\rm ej}R_\mathrm{MW}) =  \\ 
    & \hspace{0.5cm} \prod_i \int  {\rm d[Eu/Fe]} {\rm d[Fe/H]} P(d_i | {\rm [Eu/Fe]}, {\rm [Fe/H]})  \\ 
    & \hspace{1.5cm} \times P({\rm [Eu/Fe]}, {\rm [Fe/H]} | t_{\rm min},\alpha,m_{\rm ej}R_\mathrm{MW}) ,
\end{aligned}
\end{equation}
where the product is taken over all the stellar spectrum measurements and $P({\rm [Eu/Fe]}, {\rm [Fe/H]} | t_{\rm min},\alpha,m_{\rm ej}R_\mathrm{MW})$ is a delta-function likelihood located along the r-process abundance track prediction of the model with DTD parameters ($t_{\rm min}$,$\alpha$) and local BNS rate-yield $m_{\rm ej}R_\mathrm{MW}$. 

%\PL{[I notice that the code also marginalizes over a BNS enrichment efficiency parameter $f_{\rm NS}$, we should add this as well.]}\hy{not sure which parameter this is. Can you elaborate?} \PL{[It's something in the one-zone model that maps ejecta mass to r-process mass; previously it was set to 0.5, but Daniel had suggested trying to marginalize over it. I think the results are actually extremely sensitive to it and this added layer of marginalization is pretty expensive to resolve properly, so I am going to switch back to a fixed value of 0.5 and see if the plots change.]}\ds{correct, this is an efficiency factor accounting for the fact that some BNS ejecta mass may be lost due to BNS kicks (i.e. due to the fact that perhaps a significant fraction of the BNS population merges out of the Galactic plane, and only a fraction of r-process ejecta ends up in the star-forming ISM). Since we already have some generous efficiency factors in the ejecta formulae, we could as well ignore it and set it to 1 or 0.5.}\hy{Could this factor be time-dependent and thus affect our results significantly?} \PL{I would guess that it's more environment-dependent than strictly time-dependent, and allowing it to vary would make our results fuzzier overall.}

For this analysis, our prior is log-uniform for $t_{\rm min} \in [1, 2000]$ Myr and uniform for $\alpha \in [-3.0,-0.5]$. The prior on $m_{\rm ej}R_\mathrm{MW}$ is informed by the rate-yield posterior from Sec.~\ref{sec:rateyield}. In practice, we evaluate Eq.~\eqref{dtd_like} via numerical integration on a grid of points drawn along the $({\rm [Eu/Fe]}, {\rm [Fe/H]})$ track for each sample in $(t_{\rm min},\alpha,m_{\rm ej}R_\mathrm{MW})$. %Our fiducial results are conditioned on disk stars only (those from~\citet{BattistiniBensby2016}; blue diamonds in Figs.~\ref{fig:FeH_EuFe}, \ref{fig:dtd_inference_nosfr}, \ref{fig:dtd_inference}), as the one-zone chemical evolution model is strictly valid only at high metallicities when the interstellar medium has already been enriched by many r-process events and there exist well-defined average r-process abundances. However, we also investigate the effect of including the lower-metallicity halo stars by way of the SAGA database~\citep{saga}, which includes both disk and halo stars (green crosses in Figs.~\ref{fig:FeH_EuFe}, \ref{fig:dtd_inference_nosfr}, \ref{fig:dtd_inference}). 

\begin{figure}[t]
    \centering
    \includegraphics[width=1.0\linewidth]{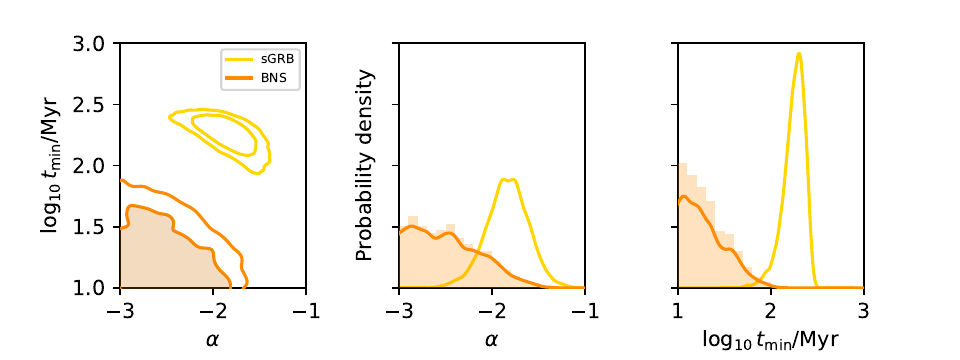}
    \includegraphics[width=1.0\linewidth, trim=0 0 0 20, clip]{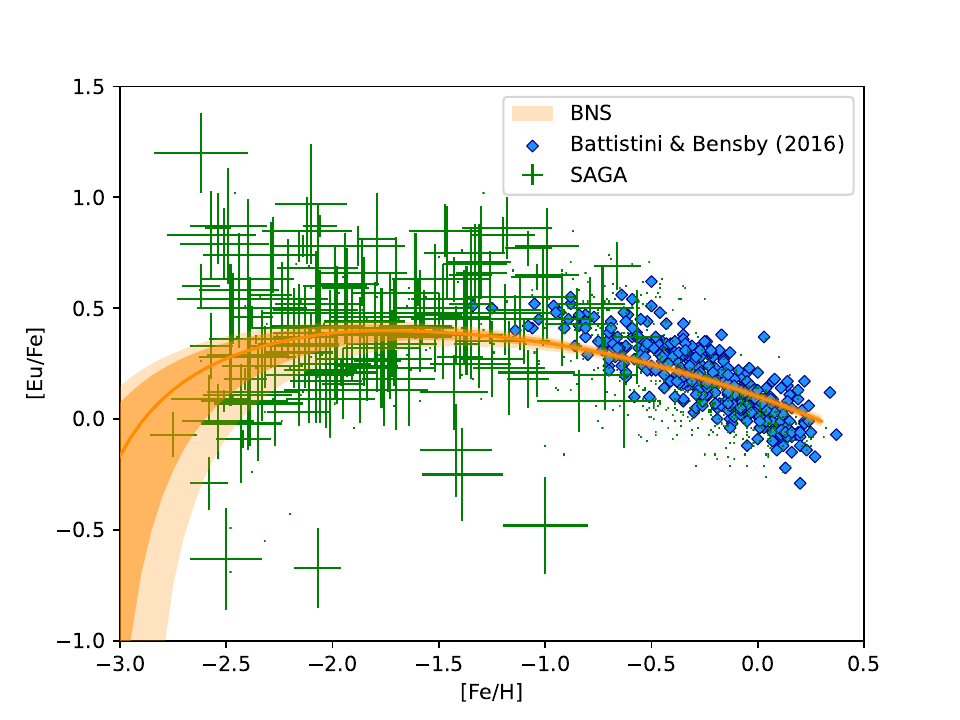}
    \caption{BNS delay time distribution parameters and [Eu/Fe] to [Fe/H] abundance ratio evolution (orange) inferred from~\citet{BattistiniBensby2016}'s disk star observations, assuming that BNS mergers are the sole site of r-process nucleosynthesis. Contours and shading represent 68\% and 90\% confidence regions. The delay time distribution parameters inferred from short gamma-ray burst observations in~\citet{ZevinNugent2022} (yellow) are shown for comparison.
    }
    \label{fig:dtd_inference_nosfr}
\end{figure}

As shown in Fig.~\ref{fig:dtd_inference_nosfr}, the disk-star data favor a very short minimum delay time, \fx{$t_{\rm min} \leq 40$ Myr} at 90\% confidence, with a steep delay time distribution, \fx{$\alpha \leq -2.0$}. This is because the disk stars' Eu abundances decreases, on average, as a function of metallicity, mirroring the decline of the cosmic star formation rate for $z \lesssim 2$ together with increased Fe dilution due to SN Ia; this decreasing trend is difficult to reproduce with long-lived BNS systems. %The problem is exacerbated by the low-metallicity halo stars, whose inclusion in the inference---ignoring inhomogeneities in the interstellar medium---requires an even shorter minimum delay time, \fx{$t_{\rm min} \leq 10$ Myr}, and steeper delay time distribution, \fx{$\alpha \leq -2.2$}.

The extremely short inferred BNS delay times are in clear tension with the sGRB-derived BNS DTD prediction of \citet{ZevinNugent2022}, which favors $t_{\rm min} \approx 180$ Myr and $\alpha \approx -1.8$~\footnote{We note that the tension also applies to the DTD values discussed in other GRB studies (e.g.,~\citet{2014MNRAS.442.2342D,wanderman_rate_2015}). However, completeness of the sample is an issue and shorter average delay times may be possible~\citep{2023A&A...680A..45S}.}. They are also inconsistent with the standard isolated binary evolution scenario for the formation of BNSs, which predicts $\alpha \approx -1.0$~\citep[e.g.,][]{Piran1992} and $t_{\rm min} \gtrsim 30$ Myr~\citep[e.g.,][]{NeijsselVignaGomez2019}. This suggests that either the BNS population has a fast-merging subpopulation, or that BNS mergers are just one channel among multiple sites for r-process nucleosynthesis. 

%\ds{How are the priors chosen? My naive expectation would be that the data would favor something close to $t_{\rm min}= 0$ (consistent with the posterior in Fig.~\ref{fig:dtd_inference}) and a very small $\alpha$ (much smaller than -1), such as to basically `remove' the DTD. We need to exclude $t_{\rm min}= 0$ from the analysis, since the time to form a NS is at least $\approx 10$\,Myr. We may also have to treat the disk stars and the halo stars separately and perhaps infer separate DTD parameters. Technically, 1-zone chemical evolution models such as the one we use here are only valid at high metallicities (disk stars) when the ISM has been enriched by many r-process events and well defined average abundances of the ISM exist. At low metallicities they are not able to capture inhomogeneities in the ISM that dominate the scatter and mean/median trends in r-process abundances.}

\section{Multi-channel r-process enrichment}\label{sec:multi}

As a way to resolve the tension between the sGRB observations and the BNS delay time distribution inferred from Galactic disk stars within the single-channel r-process model, we now assume a second r-process nucleosynthesis site that tracks the star formation history of the Milky Way but does not significantly contribute to the production of Fe. This updates the one-zone model for predicting the [Eu/Fe] vs [Fe/H] track, $P({\rm [Eu/Fe]}, {\rm [Fe/H]} | t_{\rm min},\alpha,X_{\rm SFH},m_{\rm ej}R_\mathrm{MW})$, with one more input parameter 

\begin{equation}
X_{\rm SFH} = \frac{\int m_{\rm SFH} R_{\rm SFH}(t) \,dt}{\int [m_{\rm SFH} R_{\rm SFH}(t) + m_{\rm ej} R_{\rm BNS}(t)] dt},
\end{equation}
the mass fraction of r-process abundance contributed by the second channel over Galactic history. Here, $m_{\rm SFH}$ is the per-event second-channel yield and $R_{\rm SFH}(t)$ is its rate, %per Milky Way-equivalent galaxy of the second channel, 
in parallel to $m_{\rm ej}$ and $R_{\rm BNS}(t)$ for the BNS channel.

\begin{figure}[t]
    \centering
    \includegraphics[width=1.0\linewidth]{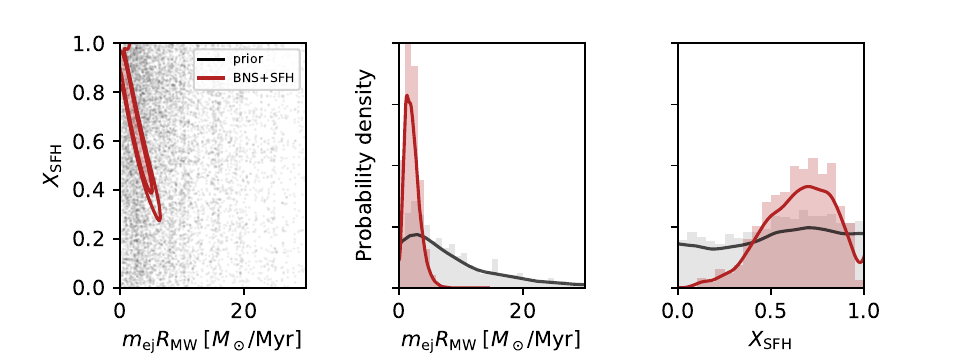}
    \includegraphics[width=1.0\linewidth, trim=0 0 0 20, clip]{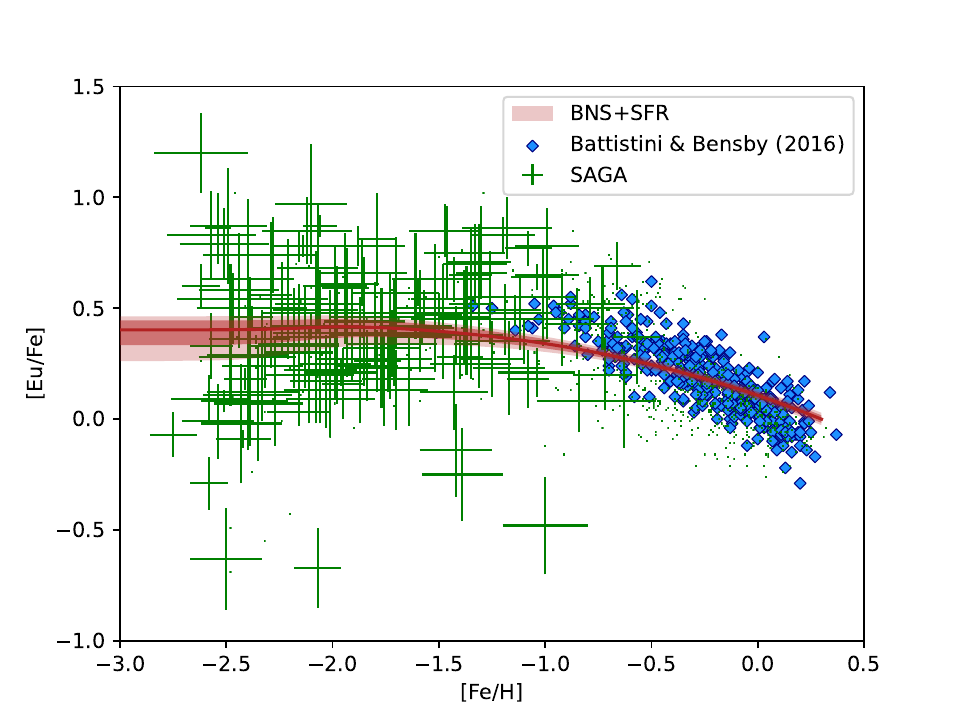}
    \caption{BNS r-process yield, second-channel contribution fraction by mass, and [Eu/Fe] to [Fe/H] abundance ratio evolution inferred from~\citet{BattistiniBensby2016}'s disk star observations, assuming that BNS mergers and a second star formation history-tracing channel contribute to r-process nucleosynthesis. The BNS delay time distribution is informed by short gamma-ray burst observations~\citep{ZevinNugent2022}. Contours and shading represent 68\% and 90\% confidence regions.}
    \label{fig:dtd_inference}
\end{figure}

We reanalyze the disk-star and SAGA observations with the sGRB-informed prior on $t_{\rm min}$ and $\alpha$, and the same prior on $m_{\rm ej} R_{\rm MW}$ as above, to constrain the second channel's contribution. We place a uniform prior on $X_{\rm SFH} \in [0, 1]$. The results are presented in Fig.~\ref{fig:dtd_inference}. The disk-star observations constrain the second channel's fractional contribution to be \fx{$X_{\rm SFH} = 0.71^{+0.20}_{-0.25}$} of the integrated r-process abundance by mass. {In other words, {45--90\%} of the r-process abundance in the Milky Way today was produced by a star-formation-tracking channel.} %This constraint is virtually unchanged if we include the halo stars in the analysis (\fx{$X_{\rm SFR} = 0.62^{+0.18}_{-0.14}$}). 
The disk-star data also tightly constrain the local BNS rate-yield to be \fx{$m_{\rm ej}R_\mathrm{MW} = 1.9^{+2.3}_{-1.4}\;M_\odot/\mathrm{Myr}$}. %(\fx{$m_{\rm ej}R_\mathrm{MW} = 2.0^{+1.2}_{-1.1}\;M_\odot/\mathrm{Myr}$} if halo stars are considered).
The Galactic stellar spectra do not constrain the BNS DTD relative to the sGRB prior within the multi-channel model. The good fit to the data in Fig.~\ref{fig:dtd_inference} demonstrates that a two-site model can reproduce Galactic stellar observations, whereas a sole-source BNS merger origin for r-process elements is inconsistent with the sGRB constraints on the DTD~\cite{ZevinNugent2022}, as well as theoretical expectations for BNS delay times, as discussed above. The data prefer the two-channel model to the BNS-only model with a Bayes factor of \fx{$\sim10^5$}, computed as a Savage-Dickey density ratio of the two-channel posterior to the prior at $X_{\rm SFH} = 0$, when the sGRB DTD is used as a prior. Remarkably, we find that the BNS merger channel and the star formation history-tracing channel \fx{have contributed at the same order of magnitude to the present-day r-process abundance in the Galaxy.}

\section{Discussion}\label{sec:discussion}

Our study of BNS mergers' contribution to r-process element enrichment in the Milky Way across cosmic time combines data from multiple kinds of astrophysical observations. We constrain the BNS merger rate evolution required to explain Galactic stellar observations, and estimate the contribution from a second r-process channel. 

Although BNS mergers' rate and per event r-process element yield are consistent with astrophysical and geological measurements, a much shorter delay relative to star formation history is required for BNS mergers alone to account for the Galactic stellar [Eu/Fe] observations at different metallicities. We therefore explore the possibility of a second r-process site with no significant time delay compared to star formation. Possible candidates for such a site include collapsars (\citet{pruet_nucleosynthesis_2003,surman_nucleosynthesis_2006,fujimoto_heavy-element_2007,SiegelBarnes2019}{; although also see~\citet{2022ApJ...934L..30J}}) or MHD supernovae \citep{nishimura_r-process_2006,winteler_magnetorotationally_2012,nishimura_intermediate_2017,halevi_r-process_2018,reichert_nucleosynthesis_2021,2021Natur.595..223Y}. The second channel's significant inferred fractional contribution of \fx{$0.71^{+0.20}_{-0.25}$} {(45--90\%)} to the present-day Galactic r-process abundance is consistent with previous estimates for a collapsar channel~\citep{SiegelBarnes2019}.

The second site could also be a separate subpopulation of BNSs. In our model, all BNS mergers follow the same rate evolution. However, Galactic observations of binary pulsars suggest that 40--60\% of BNS systems merge rapidly~\citep{Beniamini2019}, and the massive BNS merger GW190425 has been interpreted to support a shorter time from formation to merger~\citep{gw190425,RomeroShawFarrow2020,GalaudageAdamcewicz2021}. There is also theoretical support for a fast-merging BNS channel~\citep{2002ApJ...572..407B,2003MNRAS.344..629D,2003ApJ...592..475I,2023arXiv231202269B}. Moreover, there can be a subpopulation of sGRBs with shorter delay time~\citep{2024ApJ...962....5N}. Also, sGRB observations may not necessarily represent the entire BNS population, especially given recent evidence of a long GRB associated with a BNS~\citep{2022Natur.612..223R}. 

Neutron star-black hole mergers are another possible r-process site{~\citep{2023ApJ...943L..12K}}, although their contribution is highly dependent on their rate evolution and yield~\citep{2021ApJ...920L...3C}, the latter of which may be quite low based on current evidence~\citep{BiscoveanuLandry2023}. Despite the broad range of candidates for the second channel, our study places an initial constraint on the fractional contribution from this site (or combination of sites), besides typical BNS mergers.

Systematic uncertainties in the abundance data could quantitatively change our results. The assessment of the systematic uncertainty for the data we used was done by comparison with other measurements on overlapping stars. With the 0.05--0.06 dex difference in the Eu abundance (reported as $\Delta$[Eu/H] in~\citet{BattistiniBensby2016}), we do not expect a significant change in our results. Furthermore, the late-time abundance trend of [Eu/Fe] vs.~[Fe/H] is sensitive to the power-law index of the DTD of Type Ia supernovae. However, this index of $-1.1\pm0.1$ has been consistently inferred both from direct measurement using galaxy clusters \citep{freundlich_delay_2021} and from measurement of Type Ia rates and the cosmic star formation history \citep{maoz_star_2017}.

Various modeling uncertainties could also affect our results, e.g. the uncertainties in the amount of disk material ejected into disk winds in the postmerger phase (currently absent a set of long-term self-consistent simulations) and in the abundance distribution emerging from a given binary (i.e., the predicted Eu mass per event) due to uncertainties in both predicting the astrophysical conditions of the r-process and intrinsic nuclear uncertainties~\citep{horowitz_r-process_2019}. Furthermore, there are open questions in reconciling kilonova observations and other constraints on the EOS \citep{2023PhRvR...5a3168K} and in understanding the heavy element production of nuclear reaction networks under the thermodynamic conditions of the outflows \citep{2021RvMP...93a5002C}. As observational constraints tighten, this modeling uncertainty will become more significant.

More detailed models for chemical evolution and star formation histories of the Milky Way will also improve the robustness of our results. \cg{Different star formation histories could lead to horizontal shift in Figure~\ref{fig:FeH_EuFe}, and we may underestimate the uncertainties of our inference.} In particular, more detailed models would allow for the treatment of lower-metallicity halo star observations, which are impacted by inhomogeneities in the mixing of r-process elements in the ISM. Nonetheless, we do not expect these observations to alter our qualitative conclusions; even within the one-zone model, we find similar quantitative constraints on the BNS DTD, rate-yield and second-channel contribution when the halo- and disk-star SAGA observations~\citep{saga} are used in place of the~\citet{BattistiniBensby2016} disk star ones. {We have repeated our inference in the one- and two-channel scenarios with a uniform star formation rate in place of the cosmic one, to give a sense of the maximum variation in the results due to the uncertainty in the star formation history.  
We have kept the yields per event of Fe and Eu unchanged, kept the $z=0$ rate of BNS mergers consistent with the GW constraints, and normalized the rates of CCSNe and SNe Ia to the same Fe yield per stellar mass formed relative to the scenario with the cosmic star formation history. We find that the constraints on the BNS delay-time distribution and the second-channel contribution remain largely unchanged. In particular, the second-channel contribution fraction is inferred as $0.78^{+0.15}_{-0.20}$ at 90\% confidence relative to the uniform star formation history (cf.~$0.71^{+0.20}_{-0.25}$ relative to the cosmic one). This is the benefit of fitting data in the abundance space [Eu/Fe] vs. [Fe/H], which singles out the properties of Eu enrichment relative to star formation, rather than the details of the absolute star formation history. } %, which we will explore further in future work. 

{In this work, we focus on the introduction of multiple recent astrophysical constraints and the development of the Bayesian framework to properly propagate the uncertainty and quantify the contribution from different channels. In follow up work, we will use our framework to explore the impact of increasing the complexity of our modelling, by incorporating more realistic abundance yields, different star formation histories, inhomogeneous mixing of r-process elements, and variations of the DTD {for both BNSs and type Ia supernovae (see, for example~\citet{2023ApJ...943L..12K})}.}

\begin{acknowledgments}
The authors would like to thank Alexander Ji, Om Sharan Salafia, and Sharan Banagiri for useful discussions. P.L.~is supported by the Natural Sciences \& Engineering Research Council of Canada. J.S.R.~is supported by NSF PHY-2110441. We thank the Institute for Nuclear Theory at the University of Washington for its kind hospitality and stimulating research environment during INT 20r-1b. This research was supported in part by the INT's U.S. Department of Energy grant No.~DE-FG02-00ER41132. The authors are grateful for computational resources provided by the LIGO Laboratory and supported by National Science Foundation Grants PHY-0757058 and PHY-0823459. This material is based upon work supported by NSF's LIGO Laboratory which is a major facility fully funded by the National Science Foundation.

\end{acknowledgments}

\bibliography{ref}% Produces the bibliography via BibTeX.

\begin{thebibliography}{}
\expandafter\ifx\csname natexlab\endcsname\relax\def\natexlab#1{#1}\fi

\bibitem[{Abbott {et~al.}(2017)Abbott, Abbott, Abbott, \& {et
  al.}}]{abbott_gw170817_2017}
Abbott, B.~P., Abbott, R., Abbott, T.~D., \& {et al.} 2017, {\prl}, 119, 161101

\bibitem[{Abbott {et~al.}(2019)}]{LIGOScientific:2018hze}
Abbott, B.~P., {et~al.} 2019, Phys. Rev. X, 9, 011001

\bibitem[{Abbott {et~al.}(2020)}]{gw190425}
---. 2020, Astrophys. J. Lett., 892, L3

\bibitem[{Abbott {et~al.}(2023)}]{KAGRA:2021duu}
Abbott, R., {et~al.} 2023, Phys. Rev. X, 13, 011048

\bibitem[{Arcones \& Thielemann(2022)}]{arcones_origin_2022}
Arcones, A., \& Thielemann, F.-K. 2022, The Astronomy and Astrophysics Review,
  31, 1

\bibitem[{{Arnould} {et~al.}(2007){Arnould}, {Goriely}, \&
  {Takahashi}}]{Arnould2007}
{Arnould}, M., {Goriely}, S., \& {Takahashi}, K. 2007, \physrep, 450, 97

\bibitem[{Banerjee {et~al.}(2020)Banerjee, Wu, \& Yuan}]{banerjee_neutron_2020}
Banerjee, P., Wu, M.-R., \& Yuan, Z. 2020, {\apjl}, 902, L34

\bibitem[{{Battistini} \& {Bensby}(2016)}]{BattistiniBensby2016}
{Battistini}, C., \& {Bensby}, T. 2016, A\&A, 586, A49

\bibitem[{{Belczynski} {et~al.}(2002){Belczynski}, {Kalogera}, \&
  {Bulik}}]{2002ApJ...572..407B}
{Belczynski}, K., {Kalogera}, V., \& {Bulik}, T. 2002, \apj, 572, 407

\bibitem[{Beniamini {et~al.}(2016)Beniamini, Hotokezaka, \&
  Piran}]{beniamini_r-process_2016}
Beniamini, P., Hotokezaka, K., \& Piran, T. 2016, {\apj}, 832, 149

\bibitem[{{Beniamini} {et~al.}(2016){Beniamini}, {Hotokezaka}, \&
  {Piran}}]{2016ApJ...832..149B}
{Beniamini}, P., {Hotokezaka}, K., \& {Piran}, T. 2016, \apj, 832, 149

\bibitem[{Beniamini \& Piran(2019)}]{Beniamini2019}
Beniamini, P., \& Piran, T. 2019, Monthly Notices of the Royal Astronomical
  Society, 487, 4847

\bibitem[{{Beniamini} \& {Piran}(2023)}]{2023arXiv231202269B}
{Beniamini}, P., \& {Piran}, T. 2023, arXiv e-prints, arXiv:2312.02269

\bibitem[{{Biscoveanu} {et~al.}(2023){Biscoveanu}, {Landry}, \&
  {Vitale}}]{BiscoveanuLandry2023}
{Biscoveanu}, S., {Landry}, P., \& {Vitale}, S. 2023, MNRAS, 518, 5298

\bibitem[{{Capano} {et~al.}(2020){Capano}, {Tews}, {Brown}, {Margalit}, {De},
  {Kumar}, {Brown}, {Krishnan}, \& {Reddy}}]{2020NatAs...4..625C}
{Capano}, C.~D., {Tews}, I., {Brown}, S.~M., {et~al.} 2020, Nature Astronomy,
  4, 625

\bibitem[{{Chen} {et~al.}(2021){Chen}, {Vitale}, \&
  {Foucart}}]{2021ApJ...920L...3C}
{Chen}, H.-Y., {Vitale}, S., \& {Foucart}, F. 2021, \apjl, 920, L3

\bibitem[{C{\^o}t{\'e} {et~al.}(2019)C{\^o}t{\'e}, Eichler, Arcones, \& {et
  al.}}]{cote_neutron_2019}
C{\^o}t{\'e}, B., Eichler, M., Arcones, A., \& {et al.} 2019, {\apj}, 875, 106

\bibitem[{{Cowan} {et~al.}(2021){Cowan}, {Sneden}, {Lawler}, {Aprahamian},
  {Wiescher}, {Langanke}, {Mart{\'\i}nez-Pinedo}, \&
  {Thielemann}}]{2021RvMP...93a5002C}
{Cowan}, J.~J., {Sneden}, C., {Lawler}, J.~E., {et~al.} 2021, Reviews of Modern
  Physics, 93, 015002

\bibitem[{Cowan {et~al.}(2021)Cowan, Sneden, Lawler, \& {et
  al.}}]{cowan_origin_2021}
Cowan, J.~J., Sneden, C., Lawler, J.~E., \& {et al.} 2021, {\rmp}, 93, 015002

\bibitem[{{D'Avanzo} {et~al.}(2014){D'Avanzo}, {Salvaterra}, {Bernardini},
  {Nava}, {Campana}, {Covino}, {D'Elia}, {Ghirlanda}, {Ghisellini}, {Melandri},
  {Sbarufatti}, {Vergani}, \& {Tagliaferri}}]{2014MNRAS.442.2342D}
{D'Avanzo}, P., {Salvaterra}, R., {Bernardini}, M.~G., {et~al.} 2014, \mnras,
  442, 2342

\bibitem[{{Dewi} \& {Pols}(2003)}]{2003MNRAS.344..629D}
{Dewi}, J.~D.~M., \& {Pols}, O.~R. 2003, \mnras, 344, 629

\bibitem[{{Dominik} {et~al.}(2012){Dominik}, {Belczynski}, {Fryer}, {Holz},
  {Berti}, {Bulik}, {Mandel}, \& {O'Shaughnessy}}]{DominikBelczynski2012}
{Dominik}, M., {Belczynski}, K., {Fryer}, C., {et~al.} 2012, ApJ, 759, 52

\bibitem[{Drischler {et~al.}(2020)Drischler, Furnstahl, Melendez, \& {et
  al.}}]{drischler_how_2020}
Drischler, C., Furnstahl, R.~J., Melendez, J.~A., \& {et al.} 2020, {\prl},
  125, 202702

\bibitem[{Frebel \& Ji(2023)}]{frebel_observations_2023}
Frebel, A., \& Ji, A.~P. 2023, Handbook of Nuclear Physics {\textendash} Part
  III, doi:10.48550/arXiv.2302.09188

\bibitem[{Freundlich \& Maoz(2021)}]{freundlich_delay_2021}
Freundlich, J., \& Maoz, D. 2021, {\mnras}, 502, 5882

\bibitem[{Fujimoto {et~al.}(2007)Fujimoto, Hashimoto, Kotake, \& {et
  al.}}]{fujimoto_heavy-element_2007}
Fujimoto, S.-i., Hashimoto, M.-a., Kotake, K., \& {et al.} 2007, {\apj}, 656,
  382

\bibitem[{{Galaudage} {et~al.}(2021){Galaudage}, {Adamcewicz}, {Zhu},
  {Stevenson}, \& {Thrane}}]{GalaudageAdamcewicz2021}
{Galaudage}, S., {Adamcewicz}, C., {Zhu}, X.-J., {Stevenson}, S., \& {Thrane},
  E. 2021, ApJL, 909, L19

\bibitem[{{Gehrels} {et~al.}(2016){Gehrels}, {Cannizzo}, {Kanner}, {Kasliwal},
  {Nissanke}, \& {Singer}}]{2016ApJ...820..136G}
{Gehrels}, N., {Cannizzo}, J.~K., {Kanner}, J., {et~al.} 2016, \apj, 820, 136

\bibitem[{Halevi \& M{\"o}sta(2018)}]{halevi_r-process_2018}
Halevi, G., \& M{\"o}sta, P. 2018, {\mnras}, 477, 2366

\bibitem[{Horowitz {et~al.}(2019)Horowitz, Arcones, C{\^o}t{\'e}, \& {et
  al.}}]{horowitz_r-process_2019}
Horowitz, C.~J., Arcones, A., C{\^o}t{\'e}, B., \& {et al.} 2019, J. Phys. G:
  Nucl. Part. Phys., 46, 083001

\bibitem[{Hotokezaka {et~al.}(2018)Hotokezaka, Beniamini, \&
  Piran}]{hotokezaka_neutron_2018}
Hotokezaka, K., Beniamini, P., \& Piran, T. 2018, Int. J. Mod. Phys. D, 27,
  1842005

\bibitem[{Hotokezaka {et~al.}(2015)Hotokezaka, Piran, \&
  Paul}]{hotokezaka_short-lived_2015}
Hotokezaka, K., Piran, T., \& Paul, M. 2015, {\natphys}, 11, 1042

\bibitem[{{Hotokezaka} {et~al.}(2015){Hotokezaka}, {Piran}, \&
  {Paul}}]{2015NatPh..11.1042H}
{Hotokezaka}, K., {Piran}, T., \& {Paul}, M. 2015, Nature Physics, 11, 1042

\bibitem[{{Huth} {et~al.}(2022){Huth}, {Pang}, {Tews}, {Dietrich}, {Le
  F{\`e}vre}, {Schwenk}, {Trautmann}, {Agarwal}, {Bulla}, {Coughlin}, \& {Van
  Den Broeck}}]{2022Natur.606..276H}
{Huth}, S., {Pang}, P. T.~H., {Tews}, I., {et~al.} 2022, \nat, 606, 276

\bibitem[{{Ivanova} {et~al.}(2003){Ivanova}, {Belczynski}, {Kalogera}, {Rasio},
  \& {Taam}}]{2003ApJ...592..475I}
{Ivanova}, N., {Belczynski}, K., {Kalogera}, V., {Rasio}, F.~A., \& {Taam},
  R.~E. 2003, \apj, 592, 475

\bibitem[{Ji {et~al.}(2016)Ji, Frebel, Chiti, \& {et al.}}]{ji_r-process_2016}
Ji, A.~P., Frebel, A., Chiti, A., \& {et al.} 2016, {\nat}, 531, 610

\bibitem[{{Just} {et~al.}(2022){Just}, {Aloy}, {Obergaulinger}, \&
  {Nagataki}}]{2022ApJ...934L..30J}
{Just}, O., {Aloy}, M.~A., {Obergaulinger}, M., \& {Nagataki}, S. 2022, \apjl,
  934, L30

\bibitem[{{Kedia} {et~al.}(2023){Kedia}, {Ristic}, {O'Shaughnessy}, {Yelikar},
  {Wollaeger}, {Korobkin}, {Chase}, {Fryer}, \& {Fontes}}]{2023PhRvR...5a3168K}
{Kedia}, A., {Ristic}, M., {O'Shaughnessy}, R., {et~al.} 2023, Physical Review
  Research, 5, 013168

\bibitem[{Kirby {et~al.}(2020)Kirby, Duggan, {Ramirez-Ruiz}, \& {et
  al.}}]{kirby_stars_2020}
Kirby, E.~N., Duggan, G., {Ramirez-Ruiz}, E., \& {et al.} 2020, {\apjl}, 891,
  L13

\bibitem[{Kirby {et~al.}(2023)Kirby, Ji, \& Kovalev}]{kirby_r-process_2023}
Kirby, E.~N., Ji, A.~P., \& Kovalev, M. 2023, {\apj}, 958, 45

\bibitem[{{Kobayashi} {et~al.}(2020){Kobayashi}, {Karakas}, \&
  {Lugaro}}]{2020ApJ...900..179K}
{Kobayashi}, C., {Karakas}, A.~I., \& {Lugaro}, M. 2020, \apj, 900, 179

\bibitem[{{Kobayashi} {et~al.}(2023){Kobayashi}, {Mandel}, {Belczynski},
  {Goriely}, {Janka}, {Just}, {Ruiter}, {Vanbeveren}, {Kruckow}, {Briel},
  {Eldridge}, \& {Stanway}}]{2023ApJ...943L..12K}
{Kobayashi}, C., {Mandel}, I., {Belczynski}, K., {et~al.} 2023, \apjl, 943, L12

\bibitem[{{Kr{\"u}ger} \& {Foucart}(2020)}]{dyn}
{Kr{\"u}ger}, C.~J., \& {Foucart}, F. 2020, Physical Review D, 101, 103002

\bibitem[{{Legred} {et~al.}(2021){Legred}, {Chatziioannou}, {Essick}, {Han}, \&
  {Landry}}]{2021PhRvD.104f3003L}
{Legred}, I., {Chatziioannou}, K., {Essick}, R., {Han}, S., \& {Landry}, P.
  2021, Physical Review D, 104, 063003

\bibitem[{{Li} {et~al.}(2011){Li}, {Chornock}, {Leaman}, {Filippenko},
  {Poznanski}, {Wang}, {Ganeshalingam}, \& {Mannucci}}]{Li2011}
{Li}, W., {Chornock}, R., {Leaman}, J., {et~al.} 2011, \mnras, 412, 1473

\bibitem[{Lian {et~al.}(2023)Lian, Storm, Guiglion, \& {et
  al.}}]{lian_observational_2023}
Lian, J., Storm, N., Guiglion, G., \& {et al.} 2023, {\mnras}, 525, 1329

\bibitem[{{Macias} \& {Ramirez-Ruiz}(2018)}]{2018ApJ...860...89M}
{Macias}, P., \& {Ramirez-Ruiz}, E. 2018, \apj, 860, 89

\bibitem[{Macias \& {Ramirez-Ruiz}(2018)}]{macias_stringent_2018}
Macias, P., \& {Ramirez-Ruiz}, E. 2018, {\apj}, 860, 89

\bibitem[{{Madau} \& {Fragos}(2017)}]{2017ApJ...840...39M}
{Madau}, P., \& {Fragos}, T. 2017, \apj, 840, 39

\bibitem[{Maoz \& Graur(2017)}]{maoz_star_2017}
Maoz, D., \& Graur, O. 2017, {\apj}, 848, 25

\bibitem[{{Mapelli} \& {Giacobbo}(2018)}]{MapelliGiacobbo2018}
{Mapelli}, M., \& {Giacobbo}, N. 2018, MNRAS, 479, 4391

\bibitem[{{Matteucci}(2021)}]{2021A&ARv..29....5M}
{Matteucci}, F. 2021, \aapr, 29, 5

\bibitem[{{Matteucci} {et~al.}(2014){Matteucci}, {Romano}, {Arcones},
  {Korobkin}, \& {Rosswog}}]{2014MNRAS.438.2177M}
{Matteucci}, F., {Romano}, D., {Arcones}, A., {Korobkin}, O., \& {Rosswog}, S.
  2014, \mnras, 438, 2177

\bibitem[{{McMillan}(2011)}]{2011MNRAS.414.2446M}
{McMillan}, P.~J. 2011, \mnras, 414, 2446

\bibitem[{Naidu {et~al.}(2022)Naidu, Ji, Conroy, \& {et
  al.}}]{naidu_evidence_2022}
Naidu, R.~P., Ji, A.~P., Conroy, C., \& {et al.} 2022, {\apjl}, 926, L36

\bibitem[{{Neijssel} {et~al.}(2019){Neijssel}, {Vigna-G{\'o}mez}, {Stevenson},
  {Barrett}, {Gaebel}, {Broekgaarden}, {de Mink}, {Sz{\'e}csi}, {Vinciguerra},
  \& {Mandel}}]{NeijsselVignaGomez2019}
{Neijssel}, C.~J., {Vigna-G{\'o}mez}, A., {Stevenson}, S., {et~al.} 2019,
  MNRAS, 490, 3740

\bibitem[{Nishimura {et~al.}(2017)Nishimura, Sawai, Takiwaki, \& {et
  al.}}]{nishimura_intermediate_2017}
Nishimura, N., Sawai, H., Takiwaki, T., \& {et al.} 2017, {\apjl}, 836, L21

\bibitem[{Nishimura {et~al.}(2006)Nishimura, Kotake, Hashimoto, \& {et
  al.}}]{nishimura_r-process_2006}
Nishimura, S., Kotake, K., Hashimoto, M.-a., \& {et al.} 2006, {\apj}, 642, 410

\bibitem[{{Nugent} {et~al.}(2024){Nugent}, {Fong}, {Castrejon}, {Leja},
  {Zevin}, \& {Ji}}]{2024ApJ...962....5N}
{Nugent}, A.~E., {Fong}, W.-f., {Castrejon}, C., {et~al.} 2024, \apj, 962, 5

\bibitem[{{Piran}(1992)}]{Piran1992}
{Piran}, T. 1992, ApJL, 389, L45

\bibitem[{Pruet {et~al.}(2003)Pruet, Woosley, \&
  Hoffman}]{pruet_nucleosynthesis_2003}
Pruet, J., Woosley, S.~E., \& Hoffman, R.~D. 2003, {\apj}, 586, 1254

\bibitem[{{Rastinejad} {et~al.}(2022){Rastinejad}, {Gompertz}, {Levan}, {Fong},
  {Nicholl}, {Lamb}, {Malesani}, {Nugent}, {Oates}, {Tanvir}, {de Ugarte
  Postigo}, {Kilpatrick}, {Moore}, {Metzger}, {Ravasio}, {Rossi}, {Schroeder},
  {Jencson}, {Sand}, {Smith}, {Ag{\"u}{\'\i} Fern{\'a}ndez}, {Berger},
  {Blanchard}, {Chornock}, {Cobb}, {De Pasquale}, {Fynbo}, {Izzo}, {Kann},
  {Laskar}, {Marini}, {Paterson}, {Escorial}, {Sears}, \&
  {Th{\"o}ne}}]{2022Natur.612..223R}
{Rastinejad}, J.~C., {Gompertz}, B.~P., {Levan}, A.~J., {et~al.} 2022, \nat,
  612, 223

\bibitem[{Reichert {et~al.}(2021)Reichert, Obergaulinger, Eichler, \& {et
  al.}}]{reichert_nucleosynthesis_2021}
Reichert, M., Obergaulinger, M., Eichler, M., \& {et al.} 2021, {\mnras}, 501,
  5733

\bibitem[{Roederer {et~al.}(2016)Roederer, Mateo, Bailey, \& {et
  al.}}]{roederer_detailed_2016}
Roederer, I.~U., Mateo, M., Bailey, III, J.~I., \& {et al.} 2016, Monthly
  Notices of the Royal Astronomical Society, 455, 2417

\bibitem[{Roederer \& Sneden(2011)}]{roederer_heavy-element_2011}
Roederer, I.~U., \& Sneden, C. 2011, The Astronomical Journal, 142, 22

\bibitem[{{Romero-Shaw} {et~al.}(2020){Romero-Shaw}, {Farrow}, {Stevenson},
  {Thrane}, \& {Zhu}}]{RomeroShawFarrow2020}
{Romero-Shaw}, I.~M., {Farrow}, N., {Stevenson}, S., {Thrane}, E., \& {Zhu},
  X.-J. 2020, MNRAS, 496, L64

\bibitem[{{Salafia} {et~al.}(2023){Salafia}, {Ravasio}, {Ghirlanda}, \&
  {Mandel}}]{2023A&A...680A..45S}
{Salafia}, O.~S., {Ravasio}, M.~E., {Ghirlanda}, G., \& {Mandel}, I. 2023,
  \aap, 680, A45

\bibitem[{Siegel(2019)}]{siegel_gw170817_2019}
Siegel, D.~M. 2019, {\epja}, 55, 203

\bibitem[{Siegel(2022)}]{siegel_r-process_2022}
---. 2022, {\natrevphys}, 4, 306

\bibitem[{{Siegel} {et~al.}(2019){Siegel}, {Barnes}, \&
  {Metzger}}]{SiegelBarnes2019}
{Siegel}, D.~M., {Barnes}, J., \& {Metzger}, B.~D. 2019, Natur, 569, 241

\bibitem[{Sk{\'u}lad{\'o}ttir {et~al.}(2019)Sk{\'u}lad{\'o}ttir, Hansen,
  Salvadori, \& {et al.}}]{skuladottir_neutron-capture_2019}
Sk{\'u}lad{\'o}ttir, {\'A}., Hansen, C.~J., Salvadori, S., \& {et al.} 2019,
  {\aap}, 631, A171

\bibitem[{{Suda} {et~al.}(2011){Suda}, {Yamada}, {Katsuta}, {Komiya},
  {Ishizuka}, {Aoki}, \& {Fujimoto}}]{saga}
{Suda}, T., {Yamada}, S., {Katsuta}, Y., {et~al.} 2011, \mnras, 412, 843

\bibitem[{Surman {et~al.}(2006)Surman, McLaughlin, \&
  Hix}]{surman_nucleosynthesis_2006}
Surman, R., McLaughlin, G.~C., \& Hix, W.~R. 2006, {\apj}, 643, 1057

\bibitem[{Tarumi {et~al.}(2021)Tarumi, Hotokezaka, \&
  Beniamini}]{tarumi_evidence_2021}
Tarumi, Y., Hotokezaka, K., \& Beniamini, P. 2021, {\apjl}, 913, L30

\bibitem[{{The LIGO Scientific Collaboration} {et~al.}(2021){The LIGO
  Scientific Collaboration}, {The Virgo Collaboration}, {The KAGRA
  Collaboration}, \& {et al.}}]{the_ligo_scientific_collaboration_gwtc-3_2021}
{The LIGO Scientific Collaboration}, {The Virgo Collaboration}, {The KAGRA
  Collaboration}, \& {et al.} 2021, arXiv:2111.03606, arxiv:2111.03606

\bibitem[{Van Der~Swaelmen {et~al.}(2023)Van Der~Swaelmen,
  Viscasillas~V{\'a}zquez, Cescutti, \& {et al.}}]{van_der_swaelmen_gaia_2023}
Van Der~Swaelmen, M., Viscasillas~V{\'a}zquez, C., Cescutti, G., \& {et al.}
  2023, {\aap}, 670, A129

\bibitem[{Wallner {et~al.}(2015)Wallner, Faestermann, Feige, \& {et
  al.}}]{wallner_abundance_2015}
Wallner, A., Faestermann, T., Feige, J., \& {et al.} 2015, Nature
  Communications, 6, 5956

\bibitem[{Wanderman \& Piran(2015)}]{wanderman_rate_2015}
Wanderman, D., \& Piran, T. 2015, {\mnras}, 448, 3026

\bibitem[{{Wehmeyer} {et~al.}(2015){Wehmeyer}, {Pignatari}, \&
  {Thielemann}}]{2015MNRAS.452.1970W}
{Wehmeyer}, B., {Pignatari}, M., \& {Thielemann}, F.~K. 2015, \mnras, 452, 1970

\bibitem[{Winteler {et~al.}(2012)Winteler, K{\"a}ppeli, Perego, \& {et
  al.}}]{winteler_magnetorotationally_2012}
Winteler, C., K{\"a}ppeli, R., Perego, A., \& {et al.} 2012, {\apjl}, 750, L22

\bibitem[{{Yong} {et~al.}(2021){Yong}, {Kobayashi}, {Da Costa}, {Bessell},
  {Chiti}, {Frebel}, {Lind}, {Mackey}, {Nordlander}, {Asplund}, {Casey},
  {Marino}, {Murphy}, \& {Schmidt}}]{2021Natur.595..223Y}
{Yong}, D., {Kobayashi}, C., {Da Costa}, G.~S., {et~al.} 2021, \nat, 595, 223

\bibitem[{Zevin {et~al.}(2019)Zevin, Kremer, Siegel, \& {et
  al.}}]{zevin_can_2019}
Zevin, M., Kremer, K., Siegel, D.~M., \& {et al.} 2019, {\apj}, 886, 4

\bibitem[{{Zevin} {et~al.}(2022){Zevin}, {Nugent}, {Adhikari}, {Fong}, {Holz},
  \& {Kelley}}]{ZevinNugent2022}
{Zevin}, M., {Nugent}, A.~E., {Adhikari}, S., {et~al.} 2022, ApJL, 940, L18

\end{thebibliography}

\end{document}